\documentclass[conference]{IEEEtran}
\IEEEoverridecommandlockouts
\usepackage{cite}
\usepackage{amsmath,amssymb,amsfonts}
\usepackage{algorithmic}
\usepackage{graphicx}
\usepackage{textcomp}
\usepackage{xcolor}

\usepackage{pifont}
\usepackage{multirow}
\usepackage{tcolorbox}
\usepackage{booktabs}
\usepackage{mdframed}
\usepackage{adjustbox}

\def\BibTeX{{\rm B\kern-.05em{\sc i\kern-.025em b}\kern-.08em
    T\kern-.1667em\lower.7ex\hbox{E}\kern-.125emX}}

\usepackage{fancyhdr}

\fancypagestyle{firstpage}{
  \fancyhf{} % ヘッダー/フッター消去
  \fancyfoot[C]{\scriptsize © 2025 IEEE. Personal use of this material is permitted.  Permission from IEEE must be obtained for all other uses, in any current or future media, including reprinting/republishing this material for advertising or promotional purposes, creating new collective works, for resale or redistribution to servers or lists, or reuse of any copyrighted component of this work in other works.}
}

\begin{document}

\title{Development of Automated Software Design Document Review Methods Using Large Language Models\\
}

\author{\IEEEauthorblockN{Takasaburo Fukuda}
\IEEEauthorblockA{\textit{Fujitsu Ltd.} \\
Kawasaki, Japan \\
f.takasaburo@fujitsu.com}
\and
\IEEEauthorblockN{Takao Nakagawa}
\IEEEauthorblockA{\textit{Fujitsu Ltd.} \\
Kawasaki, Japan \\
nakagawa-takao@fujitsu.com}
\and
\IEEEauthorblockN{Keisuke Miyazaki}
\IEEEauthorblockA{\textit{Fujitsu Ltd.} \\
Kawasaki, Japan \\
m-keisuke@fujitsu.com}
\and
\IEEEauthorblockN{Susumu Tokumoto}
\IEEEauthorblockA{\textit{Fujitsu Ltd.} \\
Kawasaki, Japan \\
tokumoto.susumu@fujitsu.com}
}

\maketitle
\thispagestyle{firstpage}

\begin{abstract}
In this study, we explored an approach to automate the review process of software design documents by using LLM. We first analyzed the review methods of design documents and organized 11 review perspectives. Additionally, we analyzed the issues of utilizing LLMs for these 11 review perspectives and determined which perspectives can be reviewed by current general-purpose LLMs instead of humans. For the reviewable perspectives, we specifically developed new techniques to enable LLMs to comprehend complex design documents that include table data. For evaluation, we conducted experiments using GPT to assess the consistency of design items and descriptions across different design documents in the design process used in actual business operations. Our results confirmed that LLMs can be utilized to identify inconsistencies in software design documents during the review process.
\end{abstract}

\begin{IEEEkeywords}
Automated Document Review, Large Language Models, Software Design Consistency
\end{IEEEkeywords}

\section{Introduction}
The technology of LLMs has rapidly developed, beginning with the emergence of technologies such as Transformer \cite{b1} and models like BERT \cite{b2}. These advancements have led to attempts to leverage LLMs across various industries. Notably, models like the GPT series developed by OpenAI \cite{b3} have demonstrated high performance in a wide range of tasks, resulting in their increased use in both academic fields and commercial applications. In light of this background, efforts have also been made to utilize these models in the field of software engineering, particularly in the domain of code generation, where they have brought about significant transformations \cite{b4}. However, in the domain of documentation, although there are instances of using LLMs for automated text generation, such as generating software requirements specifications, code documentation, and ReadMe files \cite{b5}\cite{b6}\cite{b7}, there has been little application of LLMs in the review of software design documents.

Software design documents serve as a communication tool among stakeholders in software development, including customers and system engineers. Errors in the content of these documents can lead to significant issues such as implementation gaps and bugs in the software. To prevent such errors, frequent reviews of design documents are commonly conducted in the software development process.

However, reviewing design documents is a labor-intensive process, and errors in the design documents are often discovered in later stages of the development process, even after the reviews have been conducted \cite{b8}\cite{b9}\cite{b10}.

\begin{figure}[t]
\centerline{\includegraphics[width=\linewidth]{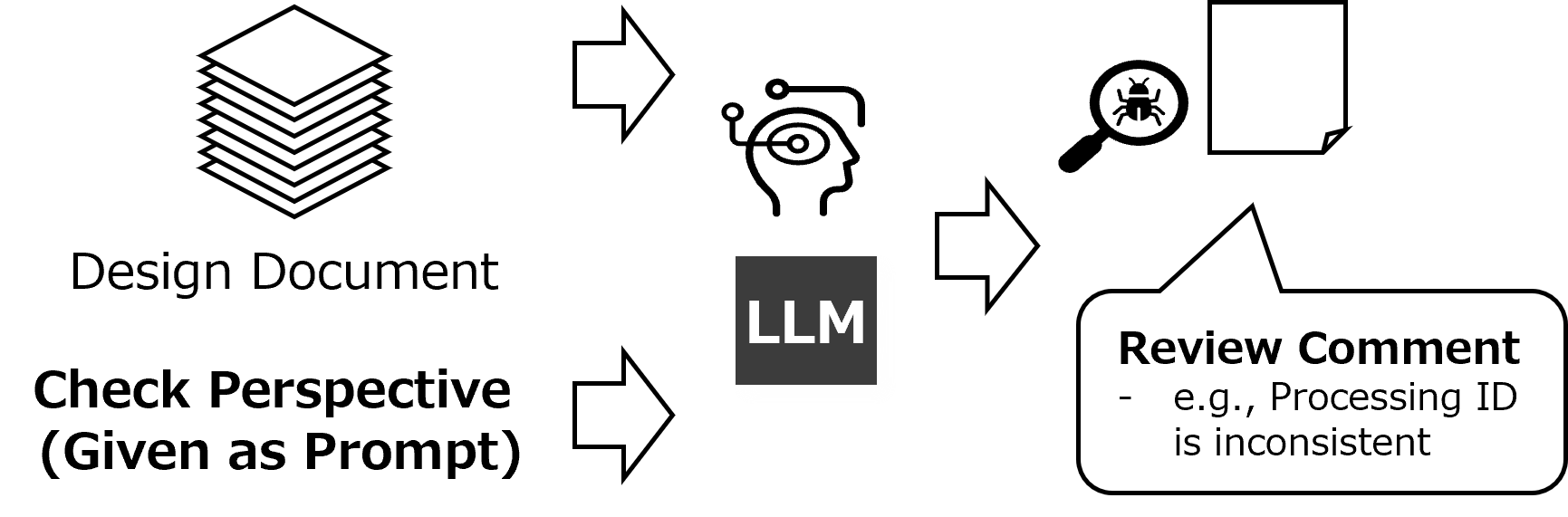}}
\caption{Illustration of LLM Utilization in the Review Process.}
\label{fig1}
\end{figure}

In this study, we propose a novel approach to automate the review process of software design documents using LLMs, with the aim of reducing review costs and improving quality. Figure \ref{fig1} illustrates the concept of utilizing LLMs in the review process.

To effectively utilize LLMs in the review of design documents, several issues need to be addressed, particularly regarding the input of detailed instructions and limitations on input tokens. Given these constraints, and to avoid the high costs associated with training specialized LLMs, we adopted a method where prompts are manually provided to the general-purpose LLM multiple times, each aligned with different review perspectives. To support this, we organized and categorized these review perspectives.

Furthermore, a unique characteristic of Japanese software design documents is that over half are created in Microsoft Excel, as it is favored for its flexibility in handling complex tabular data \cite{b11a}. This reliance on Excel introduces the need to clearly differentiate headers from values in tables—a task that LLMs often struggle to perform effectively. To address this limitation, we developed a transformation method that converts documents into a format where headers and values are distinguishable, using Markdown and JSON to enhance LLM comprehension. In evaluations conducted with real-world documents, this transformation improved defect detection accuracy by 0.43 to 0.63, highlighting the effectiveness of this approach.

Our contributions are as follows:

\begin{itemize}
    \item \textbf{Establishment of Review Perspectives}: Through discussions with expert system engineers, we identified and organized 11 review perspectives for software design documents. By categorizing these perspectives based on their suitability for LLM application, we clarified the potential scope of tasks that general-purpose LLMs could address. This categorization highlighted areas with the potential for effective utilization of general-purpose LLMs, areas requiring appropriate handling of multiple documents, and areas where the construction of specialized LLMs might be necessary.

    \item \textbf{Development of a Transformation Method for Complex Design Documents with Tabular Data}: To enable LLMs to comprehend complex design documents written in tabular formats, we developed a method that transforms these documents into a format where headers and values are distinguishable. This transformation aligns with the textual characteristics of the documents, facilitating better understanding by the LLM.

    \item \textbf{Evaluation of Automated Review Effectiveness}: We conducted evaluation experiments using actual design documents from real-world applications to assess the effectiveness of our proposed transformation method. The experiments demonstrated that our conversion method significantly improved the accuracy of the LLM's defect detection, showing practical benefits in real-world scenarios. Our findings demonstrate the practical applicability of our approach, particularly for shorter documents, while indicating the need for further research to enhance scalability for larger documents.
\end{itemize}

The remaining of this paper is organized as follows: Section 2 introduces the background of design document review methods and the utilization of LLMs in the design process. In section 3, we discuss the review perspectives and the challenges of reading tabular data with LLMs. Section 4 describes the method for reading tabular design document data with LLMs. Section 5 explains the evaluation experiments conducted using actual design documents from the field to assess the proposed method.

\section{Background and Literature Review}

\subsection{Review Techniques in Software Engineering}

Review is a crucial quality assurance technique in the development process of deliverables. General methods for reviewing software, including design documents, are described in IEEE 1028-2008 \cite{b11} and ISO/IEC 20246:2017 \cite{b12}. These methods encompass various approaches, including walkthroughs, where developers explain the deliverable to team members to gather feedback without a fixed format. Additionally, technical reviews are conducted primarily to identify issues with the deliverable, involving experts in the process. Lastly, inspections assign specific roles such as moderators and readers, and the deliverable is rigorously checked according to established rules.

IEEE 1028-2008 outlines the generic review process as follows: 1. Review Planning, 2. Overview of Procedures, 3. Preparation, 4. Examination and Evaluation, and 5. Correction and Follow-Up. During the preparation and examination phases, various reading techniques are employed to investigate the review target. Well-known reading techniques include Checklist-based Reading (CBR) \cite{b13} and Scenario-based Reading (SBR) \cite{b14}. CBR involves using a predefined checklist to review the deliverable, checking each item in a yes/no format. SBR involves identifying issues based on specific scenarios, such as use cases.

\subsection{Studies on LLM Utilization in Software Design}

According to a systematic literature review on the use of LLMs in software engineering from 2017 to 2023 \cite{b15}, only 1.29\% of the papers are related to software design. Among these, the tasks are further categorized as follows: one paper on GUI search \cite{b16}, one on specification synthesis \cite{b17}, and one on rapid prototyping \cite{b18}. Notably, there are no papers on the application of LLMs in design document reviews. Another systematic review also reported that, unlike other stages of software development, LLM-based research in the design phase is scarce \cite{b19}.

Excluding the papers in this systematic review, there are studies on automated ReadMe generation \cite{b7}, UML model generation \cite{b20}, and efforts towards bidirectional traceability between UML models and code \cite{b21}. Additionally, the MetaGPT multi-agent framework \cite{b22} generates design models by having an agent in the architect role based on requirements. Also, there are studies on the use of LLMs in UML modeling. One study\cite{b23} explores how LLMs help novice analysts create UML models. Another study\cite{b24} evaluates GPT-4v's capability to generate Java code from UML diagrams. These focus on diagrammatic representations, differing from our work, which addresses textual aspects of design documentation.

As mentioned above, the application of LLMs in the specific context of design document reviews remains underexplored. There is a significant need for efficient design document reviews, as highlighted by the quantitative data analysis of 5,546 software development projects in the financial and insurance industries \cite{b25}. This analysis reports that an average of 5.5 hours per KSLOC (thousand source lines of code) is spent reviewing design documents during the architectural design phase for new projects, and 6.4 hours per KSLOC for improvement projects. This indicates substantial labor costs and the potential for improvements through automation. Some literature suggests that practicing engineers are reluctant to rely on LLMs for higher-level design goals \cite{b19}\cite{b26}. Despite this reluctance, the demand for automating design document reviews is undeniable, given the significant time and resources currently dedicated to these tasks. Future research could address this gap, potentially leading to significant improvements in the efficiency and accuracy of design document reviews.

\section{Challenges and Approaches in Utilizing LLMs for Design Document Reviews}
\subsection{Review Perspectives and Their Classification}

\begin{center}
\begin{table*}[t]
\centering
\caption{Organization of Review Perspectives.}
\begin{tabular}{|c|c|c|c|c|c|c|}
\hline
\multirow{3}{*}{Review Perspective} & \multicolumn{2}{c|}{Difficulty Level for Non-Experts} & \multicolumn{2}{c|}{Difficulty Level for Experts} \\ \cline{2-5} 
 & Single Design Document & Multiple Design Documents & Single Design Document & Multiple Design Documents \\ \cline{2-5} 
 & Level 1 & Level 2 & Level 3 & Level 4 \\ \hline
Sufficiency Check &  & \ding{51}  &  &\\ \hline
Standards/Regulations Check &  & \ding{51} &  &\\ \hline
Traceability Check &  & \ding{51}  &  &\\ \hline
Compliance Check &  &  &  & \ding{51} \\ \hline
Functional Requirements Check & \ding{51} &  &  & \ding{51} \\ \hline
Consistency Check & \ding{51} & \ding{51} &  &  \\ \hline
Feasibility Check &  &  & \ding{51}  &  \\ \hline
Ambiguity Check & \ding{51}  &  &  &    \\ \hline
Non-Functional Requirements Check &  & \ding{51} &   & \ding{51}  \\ \hline
Cross-Sectional Check &  &  &  & \ding{51}  \\ \hline
Reflection of Comments Check &  & \ding{51}  &  &  \ding{51}  \\ \hline
\end{tabular}
\end{table*}
\end{center}

As a preliminary note, this study focuses on the review of design documents in the basic and detailed design process of software development projects. Ideally, when utilizing LLMs for design document reviews, the LLM would understand all design documents comprehensively in a single prompt input, covering all review perspectives with minimal user intervention. However, there are no specialized LLMs equipped with the knowledge and expertise comparable to seasoned system engineers necessary for making review comments. Therefore, using only the checklists typically employed in CBR aimed at "humans" is insufficient, and more detailed instructions are required.

For instance, a previous study \cite{b27} evaluating CBR review methods for UML uses a checklist example like "Are there any redundant classes?" A seasoned system engineer could determine which classes are redundant based on their knowledge and experience. However, for LLMs to perform reviews, it is necessary to clearly specify which elements of the design documents to focus on and the specific criteria for identifying redundant classes. Generally, most LLMs have limitations on the number of input tokens for prompts, thus restricting the amount of background knowledge that can be provided.

Given that LLMs with the knowledge and expertise equivalent to seasoned system engineers are not yet available, and considering the limitations on the number of input tokens, this study adopts an approach where the review process is divided into different perspectives to utilize LLMs effectively. In consultation with experts who have project management experience based on SDEM\cite{b28}, a standard system for software development created by Fujitsu Limited and compliant with the international standard ISO/IEC12207, we derived the following 11 review perspectives outlined in Appendix Table A.1, based on the following steps:

\begin{enumerate}
\item Identifying the review items that need to be checked within the design documents compliant with SDEM and related documents.
\item Defining the review items as "review perspectives" and "descriptions".
\item Classifying each review perspective based on whether it could be managed by non-experts, such as junior engineers or engineers newly assigned to the project.
\item Classifying each review perspective to determine if it requires referencing multiple documents.
\end{enumerate}

Based on the steps outlined above, the 11 review perspectives are summarized in Table 1. In Table 1, "Difficulty Level for Non-Experts" refers to the level of difficulty that allows reviews to be conducted by junior SEs without specialized knowledge or by third parties outside the project developing the design documents. Additionally, perspectives that require referencing multiple documents during the review are categorized as "Multiple Design Documents," while those that do not require such references are categorized as "Single Design Document." Consequently, review perspectives marked on the right side of the table indicate higher difficulty levels, with levels set from 1 to 4 for each stage.

Based on Table 1, we consider the applicability of LLMs for each review perspective. To utilize LLMs for perspectives corresponding to Levels 3 and 4, it is essential for the LLMs to possess knowledge comparable to that of experienced system engineers. For instance, the feasibility check requires verifying whether implementation and maintenance are possible based on the design documents, necessitating a review by individuals with specialized knowledge. Therefore, for review perspectives at these levels, it is necessary to develop specialized LLMs trained on a combination of general software engineering knowledge, domain-specific knowledge of the system being implemented, and the unique knowledge frameworks specific to the development environment. Currently, general-purpose LLMs widely available do not possess domain-specific knowledge required for system implementation. Although many of these models have been trained on general software engineering standards such as ISO/IEC, they lack the specialized knowledge necessary for higher-level review tasks. Therefore, this study focuses on Levels 1 and 2, which can be managed by general-purpose LLMs.

\begin{figure}[t]
\centerline{\includegraphics[width=\linewidth]{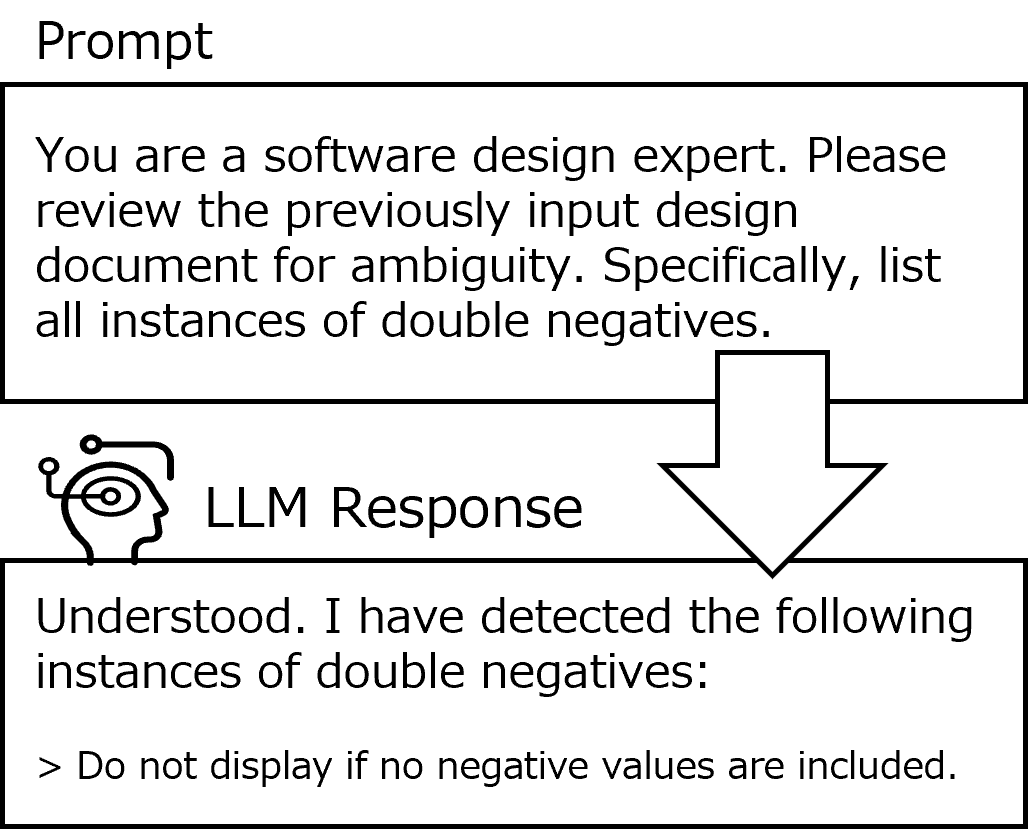}}
\caption{Example of Usage in Ambiguity Check.}
\label{fig2}
\end{figure}

Next, we consider the perspectives corresponding to Level 2. These perspectives require referencing multiple appropriate design documents to conduct the review. For instance, in the traceability check, it is necessary to confirm that the content defined in the upstream design documents is not omitted in the downstream design documents. This requires selecting and referencing multiple related design documents accurately. Existing research proposes techniques to automatically identify related documents based on design and workflow information \cite{b9}\cite{b29}, and these can be combined in practical operations. In this study, we consider a method where LLMs are utilized for review perspectives at Levels 1 and 2 by manually selecting and inputting the design documents to be referenced. For example, in the ambiguity check, to ensure that the expressions in the design documents are clear, specific items to be checked, such as "avoid using double negatives" as shown in Figure \ref{fig2}, are provided as prompts along with the design document descriptions. The LLM is then expected to output responses based on these perspectives.

\subsection{Challenges in Reviewing Complex Design Documents}

\begin{figure}[h]
\centerline{\includegraphics[width=\linewidth]{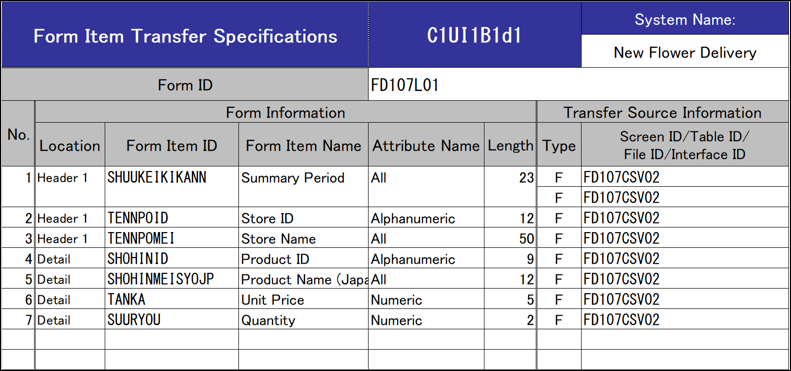}}
\caption{Example of a Design Document in the Architectural Design Phase of SDEM.}
\label{fig3}
\end{figure}

Considering that the design documents to be input into the LLM for the desired review perspectives have been manually selected, we next consider the method of loading these design documents into the LLM. Generally, in Japanese systems integrator, software design documents are often created using spreadsheet software such as Microsoft Excel \cite{b11a}. These design documents frequently use complex table structures, as shown in Figure \ref{fig3}. In fact, in the aforementioned SDEM, 41 out of the 47 standard design documents in the architectural design phase use table formats. There are also design documents that use images for screen layouts and process flows. However, reading such documents would require additional multimodal support, which is beyond the scope of this study.

\begin{figure}[h]
\centerline{\includegraphics[width=\linewidth]{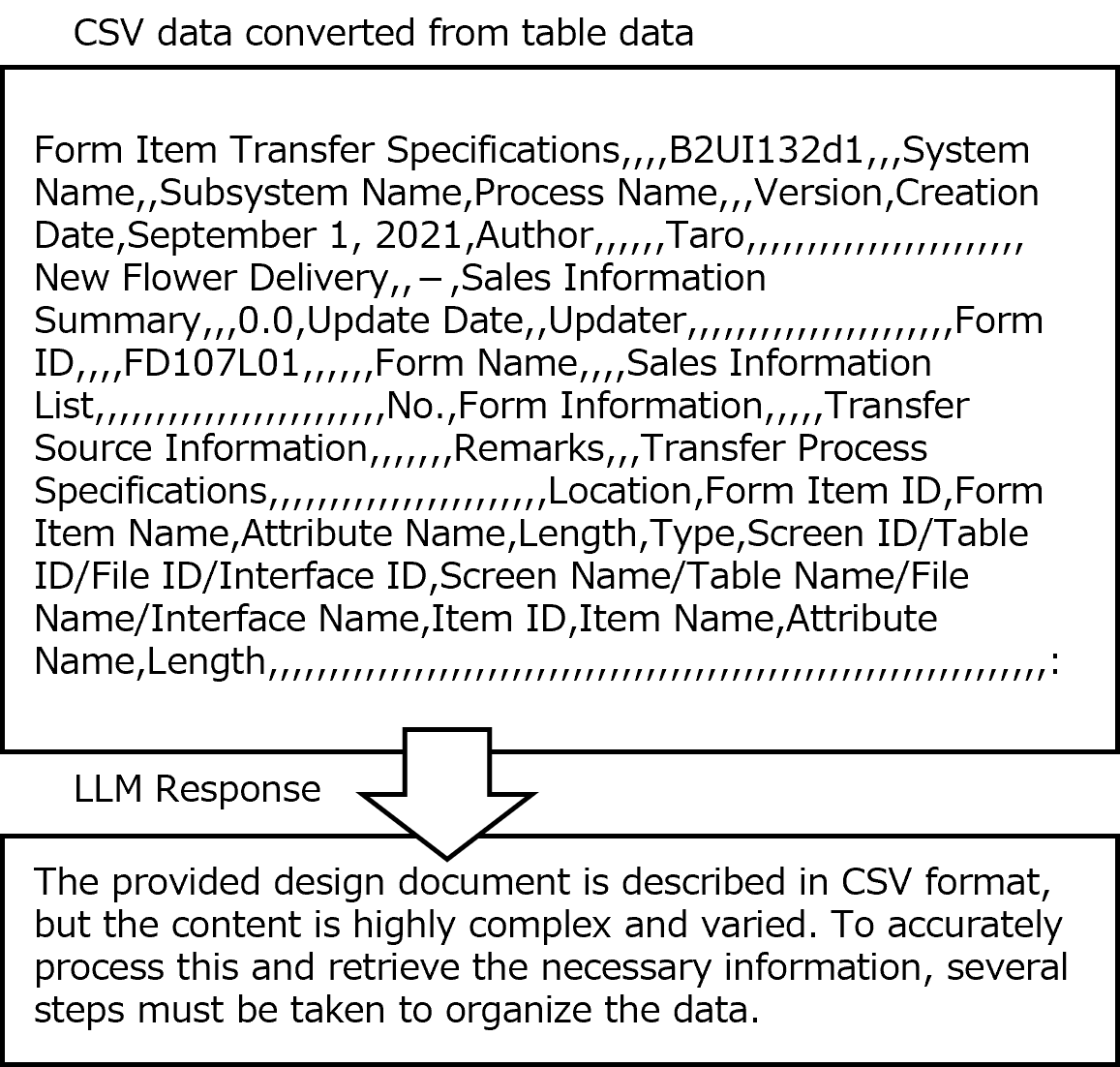}}
\caption{Example of Converting a Complex Table Structure into CSV Format and Inputting it into an LLM for Review Responses}
\label{fig4}
\end{figure}

When loading design documents with table structures like those in Figure 3 into an LLM, it is common to first convert the table into CSV format before inputting it into the LLM. Although there is existing research on enabling LLMs to understand CSV data, it is well known that there are still significant challenges in semantic analysis and response generation \cite{b30}. Especially in the case of software design documents, which frequently have complex table structures like those in Figure 3, it is often difficult for LLMs to recognize CSV data accurately. For instance, if the design document shown in Figure 3 is converted to CSV and then directly input into the LLM, as shown in Figure \ref{fig4}, the LLM often fails to understand the CSV content, rendering it unable to perform the review. Therefore, to effectively utilize LLMs in the review of design documents, it is essential to address such complex table structures.

\section{Proposed Method}

\subsection{Data Conversion to Distinguish Headers and Values}

In order to effectively utilize LLMs for the review of software design documents without resorting to costly fine-tuning or specialized LLMs, it is crucial to enable general-purpose LLMs to understand the relationships between headers and values within complex table structures. Prior research has emphasized the importance of recognizing these relationships for accurate comprehension \cite{b31}.

In preliminary tests, we attempted to improve LLM inference accuracy by appending header information to CSV data, using layout cues within the table structure to estimate which elements appeared to be headers and which appeared to be values. Despite these efforts, simply adding such information did not sufficiently enhance LLM performance. This study identified that ``the CSV data used to train the LLM does not inherently include information to distinguish between headers and values.’’ When making inferences based on such data, the LLM may proceed with inferences without understanding which elements are headers and which are values when the data remains in CSV format. Therefore, this study proposes a method where, before inputting data described in a table structure into the LLM, it is converted into a format that can distinguish between headers and values.

As a specific example of a format that can distinguish between headers and values, consider Markdown format. Unlike CSV data, Markdown data includes header symbols that make it obvious during the training data stage which elements are headers and which are values. By converting the data into Markdown format before inputting it, the LLM can make inferences based on training data where the relationship between headers and values is clear, like in Markdown format. This approach is expected to allow the LLM to understand and infer the relationship between headers and values more accurately.

\begin{figure}[h]
\centerline{\includegraphics[width=\linewidth]{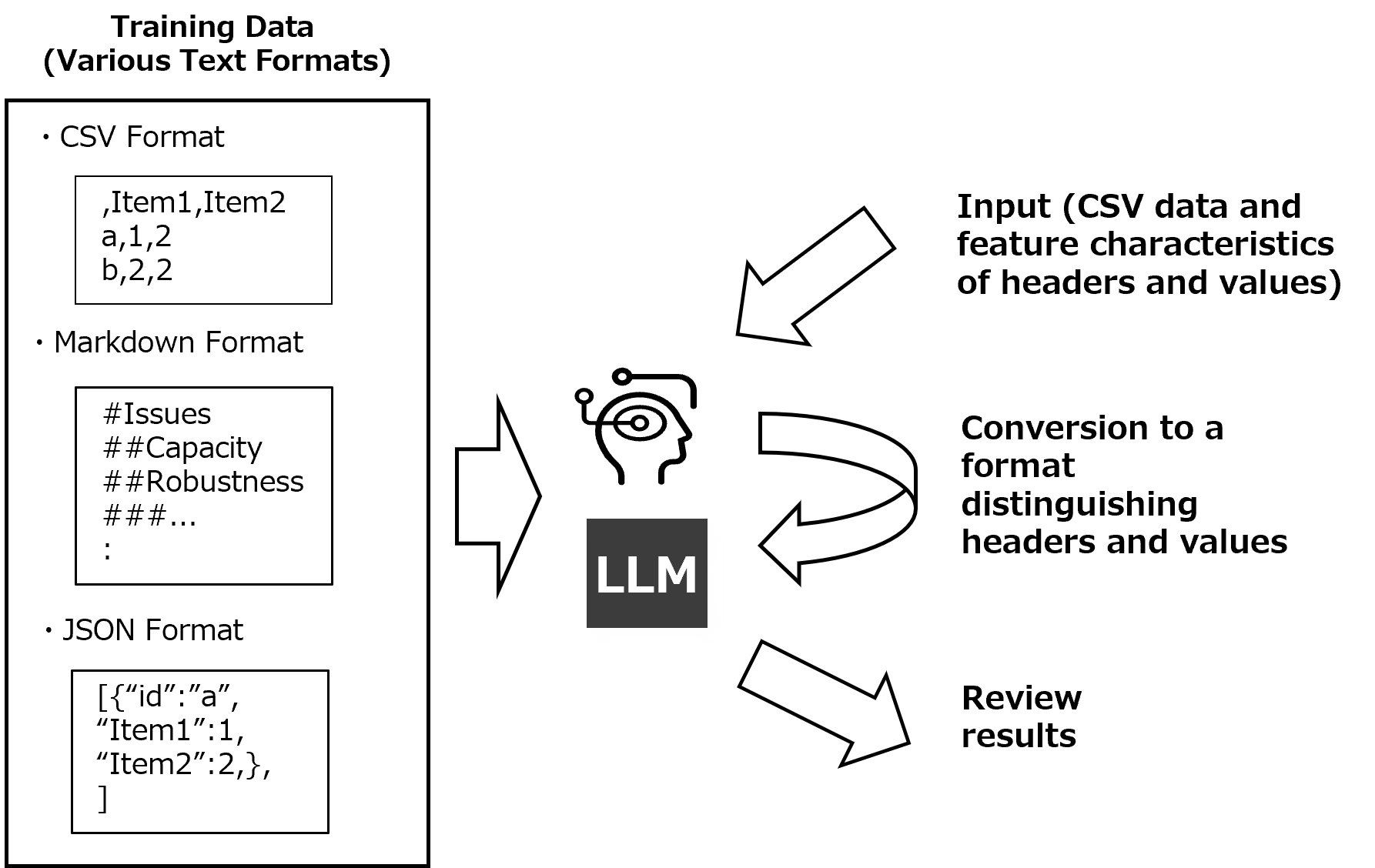}}
\caption{Overview of the Proposed Method: Converting CSV Input to a Format Distinguishing Headers and Values, and Generating Review Results.}
\label{fig5}
\end{figure}

The overview of the proposed method is shown in Figure \ref{fig5}. It is assumed that the conversion to a format where headers and values can be distinguished, such as Markdown format, will be performed using an LLM. The reason for using an LLM for this conversion is to follow the concept of the Chain-of-Thought (CoT) prompting technique \cite{b32}, which aims to incrementally make the LLM understand the headers and values during the inference process of the review.

\subsection{Selection of Conversion Formats}

In addition to Markdown format, other formats such as JSON can also be used to distinguish between headers and values. We hypothesize that selecting these conversion formats based on the characteristics of the text might yield better results. For instance, Markdown format is often used for natural language descriptions and might therefore contribute to inference accuracy in data containing or resembling natural language constructs. On the other hand, JSON format is frequently used for symbolic representations in programming, which suggests that it could enhance inference accuracy in data rich in symbolic information.

To select the appropriate conversion format, it is conceivable to construct a classifier that learns the characteristic features of documents commonly used in each format, based on training data similar to the domain of the design documents to be reviewed. However, collecting Markdown or JSON documents relevant to the specific domain can be resource-intensive. Therefore, as a more practical and simpler approach that does not incur high collection costs, we propose using part-of-speech (POS) information for classification. This method is designed to be a pragmatic solution and is specifically intended to classify data into two formats: Markdown and JSON.

Specifically, morphological analysis is performed on all elements of the data, and the proportion of lexical items tagged as ``symbols'', ``characters'', and ``nouns (without dictionary information)'' is used to determine the conversion format. For instance, if the proportion of ``symbols'' is high, the data is considered rich in symbolic information and is converted to JSON format. Conversely, if the proportion is low, indicating a predominance of natural language information, the data is converted to Markdown format.

To formalize this approach, let \( P_s \) represent the proportion of symbols, characters, and nouns (without dictionary information) in the text. The decision rule for selecting the format can be expressed as follows:

\begin{equation}
\text{Format} = 
\begin{cases} 
\text{JSON} & \text{if } P_s \geq \theta_s \\
\text{Markdown} & \text{if } P_s < \theta_s 
\end{cases}
\end{equation}

Where \( \theta_s \) is a predefined threshold value for the proportion of these elements. This threshold can be empirically determined based on the characteristics of the training data. By implementing this decision rule, we ensure that the conversion format is selected in a manner that is both efficient and effective, leveraging the inherent linguistic features of the input data. In determining  \( \theta_s \) for distinguishing between formats, we applied a methodology inspired by techniques commonly used in binary classification tasks, particularly those designed to handle imbalanced datasets. We first analyzed the distribution of symbolic and natural language elements within a representative set of design documents, then used metrics such as the F1 score to balance precision and recall for format classification. By selecting the threshold that maximized the F1 score, we ensured that the chosen value effectively differentiates between documents suited for Markdown and those better aligned with JSON.

While this approach effectively determines \( \theta_s \) for distinguishing between formats, implementing it rigorously requires labeling each document in terms of whether it predominantly contains natural language or symbolic content. This project-specific labeling process can be labor-intensive. In practical operations, we streamline this by initially setting parameters based on past experience and typical document characteristics observed in similar projects. If operational issues arise, manual adjustments to the threshold are performed as needed. This approach leverages experience to reduce the burden of extensive labeling while maintaining flexibility to fine-tune the system during real-world application.

\section{Evaluating LLM-Assisted Review Processes}
\subsection{Research Questions}

In this section, we address the following research questions (RQ1, RQ2, and RQ3) to evaluate the effectiveness of converting design documents with table structures into a format that can distinguish between headers and values for LLM comprehension, the effectiveness of automatically selecting conversion formats based on data characteristics, and the scalability of LLM-assisted reviews in real-world work environments.

\begin{itemize}
    \item \textbf{RQ1}: Is converting design documents into a format that distinguishes between headers and values effective in improving defect detection performance using LLMs?
    \item \textbf{RQ2}: Is the selection of conversion formats based on data characteristics effective in improving defect detection performance using LLMs?
    \item \textbf{RQ3}: Does the scalability of reviews utilizing LLMs meet the practical levels required in actual work environments, considering the volume of text?
\end{itemize}

\subsection{Evaluation Setup: Experimental Data and LLM Environment}

To conduct the evaluation experiments, we focused on the consistency check among the 11 review perspectives listed in Table 1. This focus was motivated by two main reasons. First, among the Level 1 and Level 2 perspectives that are feasible with general-purpose LLMs, consistency checks are particularly sensitive to the LLM’s ability to correctly interpret headers and values within complex table structures. Unlike ambiguity checks, where each sentence can be reviewed individually without relying heavily on table structure, consistency checks require an accurate understanding of relationships across headers and values in multiple sections. When the LLM misinterprets these elements, it risks failing to accurately assess consistency across documents, as the recognition of structured data relationships is crucial. Second, consistency checks represent one of the most frequent and essential review tasks in typical design document reviews, as inconsistencies in terminology and process identifiers across documents are a common type of defect \cite{b33}.

Within the consistency check perspective, we particularly emphasized the "bidirectional" relationship, which involves verifying inconsistencies in both directions between corresponding design documents \cite{b10}. For example, if a specific process ID is labeled differently in two related design documents, it is flagged as a bidirectional inconsistency.

In these experiments, we prepared pairs of design documents to perform bidirectional checks, with examples of the original and modified design documents provided in Appendix (Figure A.1 and Figure A.2).

Figure A.1 in the appendix shows an actual Excel-based design document used for the consistency checks in this study. The document adheres to the SDEM standard and was derived from a banking account system design. Sensitive information such as system names and developer names has been replaced with dummy data. In Figure A.2, an intentional inconsistency was introduced to test the LLM’s ability to detect errors. For example, the process ID "execute" in one document was changed to "execution" in the modified version, representing a bidirectional inconsistency that the LLM was tasked to identify. The Process ID is essential for ensuring traceability across design documents, as it links related processes throughout the documentation. 

Additionally, we included other common types of errors—such as differences in ID numbers, text labels, preconditions, and variable types—in various parts of the documents. These types of inconsistencies were selected because they frequently impact traceability and accuracy in real-world design documents. For instance, discrepancies in IDs and text labels can hinder traceability by breaking links across documents, while mismatches in preconditions or variable types can lead to implementation issues. 

In conducting evaluations using LLMs, it is crucial to ensure that the LLMs used do not pose any security risks, especially when using design documents employed in actual business operations. Therefore, in this study, we contracted with Microsoft Azure as a company and performed all evaluations exclusively within the GPT environment provided via Azure. For the evaluations, we used three different GPT models: gpt-35-turbo (1106), gpt-4 (turbo-2024-04-09), and gpt-4o (2024-05-13). Throughout all experiments, we set the parameters to their default values for both temperature and top-p to maintain consistency and reproducibility across models, ensuring that any observed differences in performance could be attributed to the models themselves rather than varied parameter settings.

When inputting the design documents into the LLM, we first convert the data into CSV format and estimate the headers and values using the method described in the previous section. This information is then added to the data. During inference, we instruct the LLM to convert the data into a format that can distinguish between headers and values, as described in the previous section. After conversion, we provide prompts for conducting consistency checks and let the LLM perform the review. Below is an example of the prompt used for consistency checks.

\begin{center}
\textbf{Prompt for Consistency Checks:}
\end{center}

\textbf{[Request]} Based on the two input design documents, please conduct a review from the perspective of consistency.

\textbf{[Output Format]}
\begin{itemize}
\item Perspective:
\item Presence of Inconsistencies: (Yes or No)
\item Inconsistent Locations:
\item Suggested Corrections:
\item Justification:
\end{itemize}

\textbf{[Supplementary Notes]}
\begin{itemize}
\item If no inconsistencies are found, state "Presence of Inconsistencies: No".
\item There may be multiple points of inconsistency.
\end{itemize}

In this experiment, we used a straightforward prompt that only instructs the LLM to conduct consistency checks, as shown above. Due to this general approach, the LLM may occasionally detect inconsistencies beyond the intentionally introduced errors. To ensure that our evaluation focuses on relevant inconsistencies, we established specific criteria for assessing the detected discrepancies.

For calculating precision, we applied the following standards to classify detected inconsistencies:

\begin{itemize} 
\item Errors that are generally acceptable by convention but could appear inconsistent from a certain perspective are not counted as inconsistencies in our data. 
\item Obvious incorrect detections—those that clearly do not reflect any inconsistency within the design context—are marked as false positives. 
\end{itemize}

\subsection{Experimental Results for RQ1}

To evaluate RQ1, we conducted experiments comparing the review results based on data in its original CSV format and data converted to Markdown format before review. We used pairs of design documents (A and B) similar to those in Figure A.1 and Figure A.2, specifically selecting documents with a high natural language content. According to Equation (1), these documents had a \( P_s \) value below the predefined threshold, indicating a predominance of natural language over symbolic elements, making them suitable for Markdown conversion. In each document B, we intentionally introduced two defects. Each design document used was small in scale, with fewer than 500 characters in Japanese. We prepared five different unique combinations of design documents A and B, resulting in five distinct document pairs. For each pair, we conducted 10 experimental runs, totaling 50 runs. In this experiment, we evaluated both the presicion and recall of each GPT model in identifying the introduced defects. The evaluation results are shown in Table 2.

\begin{table*}[h]
\caption{Evaluation Results for RQ1: Consistency Check for CSV and Markdown Formats}
\centering
\begin{tabular}{lcccccc}
\toprule
 & \multicolumn{2}{c}{\textbf{gpt-35-turbo}} & \multicolumn{2}{c}{\textbf{gpt-4}} & \multicolumn{2}{c}{\textbf{gpt-4o}} \\
\cmidrule(lr){2-3} \cmidrule(lr){4-5} \cmidrule(lr){6-7}
 & Precision & Recall & Precision & Recall & Precision & Recall \\
\midrule
\textbf{Unconverted Method} & 0.93 & 0.25 & 0.91 & 0.33 & 0.94 & 0.53 \\
\textbf{Proposed Method} & 0.96 & 0.87 & 0.98 & 0.96 & 0.99 & 0.96 \\
\bottomrule
\end{tabular}
\label{tab:evaluation-results}
\end{table*}

As a result of the experiments, significant differences were observed between cases where the review was conducted without any conversion and cases where documents were converted to Markdown format using the proposed method. While Precision remained relatively stable across formats, Recall improved substantially in the Markdown format. This improvement can be attributed to the clearer distinction between headers and values, making elements for consistency checks more explicit. Figure 6 shows the design document from Figure A.1 after conversion to Markdown, illustrating how this format enables the LLM to better differentiate headers and values, thus enhancing its accuracy in detecting relevant inconsistencies. These results demonstrate the effectiveness of the proposed method, highlighting the importance of proper data formatting in improving LLM performance, particularly for Recall in consistency checks.

\begin{figure}[h]
\centerline{\includegraphics[width=\linewidth]{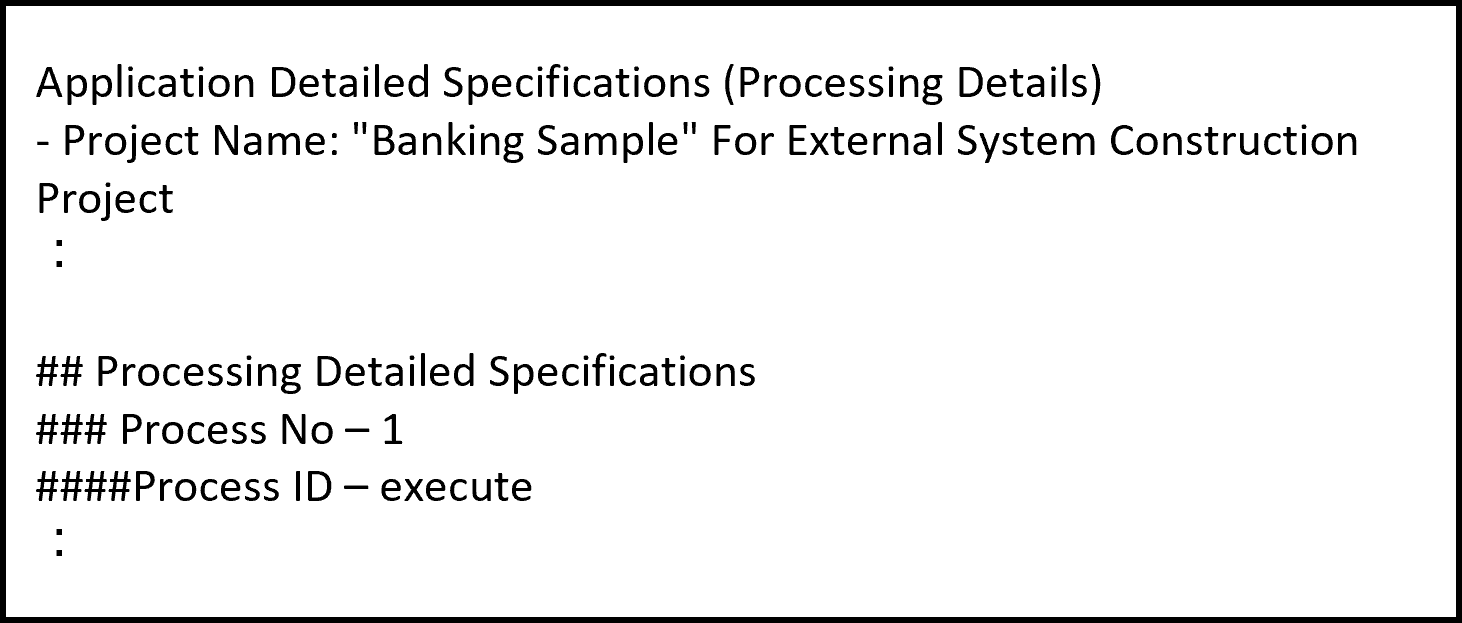}}
\caption{Example of Converted Design Document in Markdown Format for Consistency Check.}
\label{fig7}
\end{figure}

In cases where the proposed method did not perform well, it was observed that the gpt-35-turbo model sometimes concluded with a general explanation of consistency review. Additionally, all models occasionally only partially converted the data into Markdown format. We believe that these issues can be mitigated to some extent through improved prompting techniques, such as providing more detailed instructions and examples or using intermediate prompts to guide the conversion process more effectively. 

\begin{tcolorbox}[colframe=black,colback=white,boxrule=0.5mm]
\textbf{Answer to RQ1:} Based on the experimental results, the proposed method significantly improves defect detection performance compared to the unconverted method. Specifically, recall increased by a range of 0.43 to 0.63 across different models. This demonstrates the effectiveness of converting design documents to a format that clearly distinguishes between headers and values for LLM-based reviews.
\end{tcolorbox}

\begin{table*}[t]
\caption{Evaluation Results for RQ2: Precision and Recall for Natural Language-Rich Documents}
\centering
\begin{tabular}{lcccccc}
\toprule
 & \multicolumn{2}{c}{\textbf{gpt-35-turbo}} & \multicolumn{2}{c}{\textbf{gpt-4}} & \multicolumn{2}{c}{\textbf{gpt-4o}} \\
 & \textbf{Precision} & \textbf{Recall} & \textbf{Precision} & \textbf{Recall} & \textbf{Precision} & \textbf{Recall} \\
\midrule
\textbf{Markdown Format} & 0.96 & 0.87 & 0.98 & 0.96 & 0.99 & 0.96 \\
\textbf{JSON Format} & 0.88 & 0.47 & 0.92 & 0.67 & 0.93 & 0.94 \\
\bottomrule
\end{tabular}
\label{tab:natural-language}
\end{table*}

\begin{table*}[t]
\caption{Evaluation Results for RQ2: Precision and Recall for Symbolic Representation-Rich Documents}
\centering
\begin{tabular}{lcccccc}
\toprule
 & \multicolumn{2}{c}{\textbf{gpt-35-turbo}} & \multicolumn{2}{c}{\textbf{gpt-4}} & \multicolumn{2}{c}{\textbf{gpt-4o}} \\
 & \textbf{Precision} & \textbf{Recall} & \textbf{Precision} & \textbf{Recall} & \textbf{Precision} & \textbf{Recall} \\
\midrule
\textbf{Markdown Format} & 0.89 & 0.38 & 0.90 & 0.57 & 0.92 & 0.71 \\
\textbf{JSON Format} & 0.93 & 0.90 & 0.95 & 0.98 & 0.96 & 0.95 \\
\bottomrule
\end{tabular}
\label{tab:symbolic-representation}
\end{table*}

\subsection{Experimental Results for RQ2}

To evaluate RQ2, we first converted the design documents used in Experiment 1 into JSON format instead of Markdown format. Additionally, we prepared an alternative set of design documents with a high frequency of symbolic representations, as shown in Figure A.3. These documents were specifically selected based on Equation (1) to have \( P_s \) value above the predefined threshold, indicating a predominance of symbolic content. For these symbol-rich documents, we evaluated the performance of the models when converted to both Markdown and JSON formats.

Each design document used in the experiments contained fewer than 500 characters in Japanese. We created five distinct patterns for these document pairs, with two intentional errors introduced in each document B, similar to Experiment 1. For each of the five patterns, we conducted 10 experimental runs, resulting in a total of 50 runs. The evaluation results are shown in Table 3 and Table 4.

The experimental results indicated that, for gpt-35-turbo and gpt-4, data with a high natural language content had higher detection rates when converted to Markdown format. However, for gpt-4o, the detection rate remained the same regardless of the format. For data with a high frequency of symbolic representations, all models achieved higher detection rates when converted to JSON format. When natural language-rich data was converted to JSON format, frequent occurrences of nested structure misalignments and missed headers and values, as shown in Figure 7, were observed with gpt-35-turbo and gpt-4. In contrast, such phenomena were rarely seen with gpt-4o, indicating that it can more intelligently handle and convert natural language-rich descriptions to JSON format compared to the previous models.

\begin{figure}[t]
\centerline{\includegraphics[width=\linewidth]{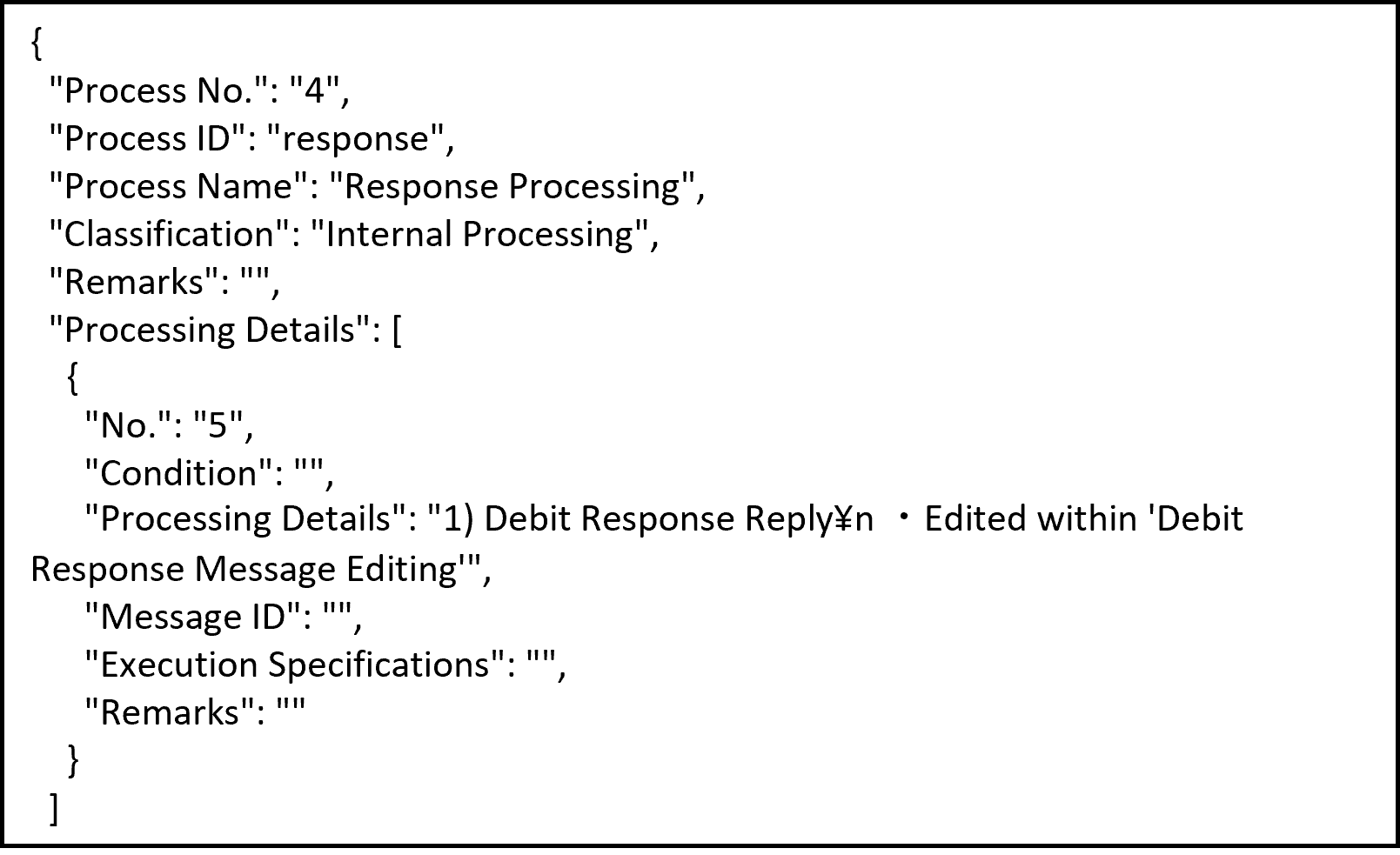}}
\caption{Example of Nested Structure Misalignments and Missed Headers and Values in JSON Format for Natural Language-Rich Document.}
\label{fig9}
\end{figure}

For data with a high frequency of symbolic representations, when converted to JSON format, the headers and values were mostly correctly distinguished, as shown in Figure 8. However, when converted to Markdown format, there were many cases where a simple pipe expression was used, as shown in Figure 9. This phenomenon occurred across all models and led to a decrease in detection rates when converted to Markdown format. These results suggest the effectiveness of selecting the conversion format based on the characteristics of the data.

\begin{figure}[t]
\centerline{\includegraphics[width=\linewidth]{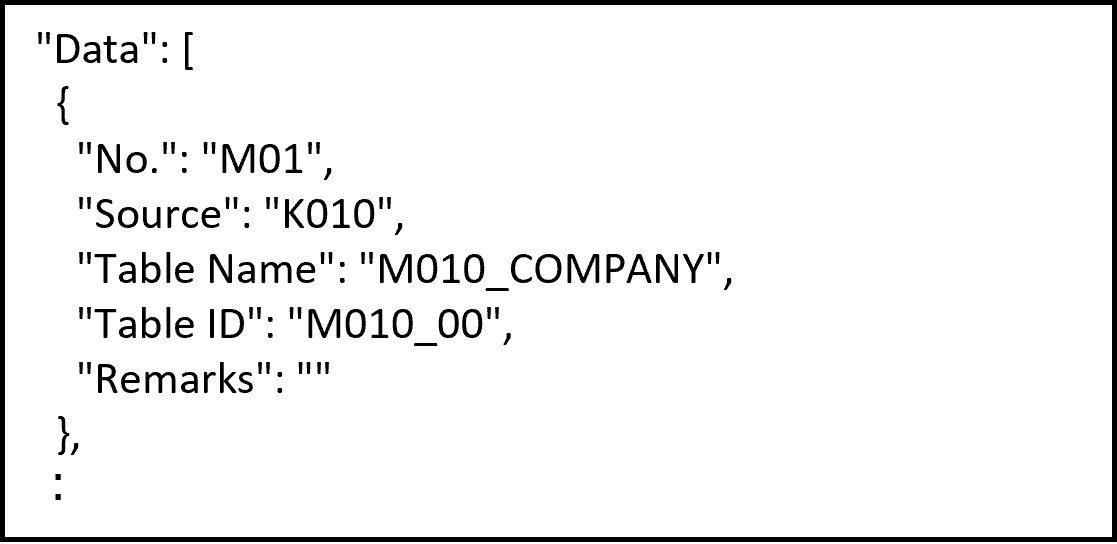}}
\caption{Example of Correct Header and Value Distinction in JSON Format for Symbolic Representation-Rich Data.}
\label{fig10}
\end{figure}

\begin{figure}[t]
\centerline{\includegraphics[width=\linewidth]{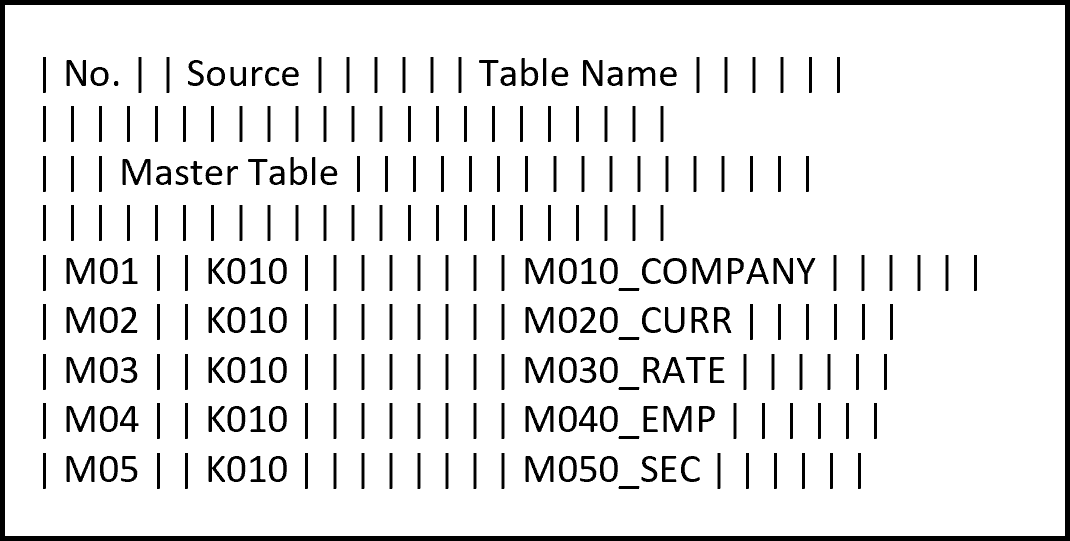}}
\caption{Example of Simple Pipe Expression Conversion in Markdown Format for Symbolic Representation-Rich Data.}
\label{fig10}
\end{figure}

\begin{tcolorbox}[colframe=black,colback=white,boxrule=0.5mm]
\textbf{Answer to RQ2:} The experimental results indicate that detection rates improve when data is converted to the format best suited to its characteristics. Natural language-rich data performs better in Markdown format, while symbolic representation-rich data performs better in JSON format. This confirms the effectiveness of selecting the appropriate conversion format based on data characteristics.
\end{tcolorbox}

\subsection{Experimental Results for RQ3}

In previous experiments, we focused on design documents with fewer characters to confirm the effectiveness of the proposed method. This time, to assess whether the method can be applied in real-world operations, we evaluated its scalability with respect to document length. As the number of characters increases, it is anticipated that consistency checks will become more challenging. For RQ3, we conducted experiments using design documents with a high amount of natural language content, consistently converting them into Markdown format. All character counts were measured in Japanese. We prepared multiple documents of varying lengths: fewer than 500 characters (from previous experiments), 500-1500 characters, 1500-2500 characters, 2500-4000 characters, 4000-5000 characters, 5000-6000 characters, and 6000-7000 characters.

Based on the results from RQ1 and RQ2, where Precision showed minimal variation across different conditions, this experiment focused solely on evaluating Recall. The method of introducing errors and the number of design documents followed the same protocol as in previous experiments. The evaluation results are shown in Table 5.

\begin{table}[h]
\caption{Detection Rates for Various Document Lengths}
\centering
\begin{tabular}{lcccc}
\toprule
\textbf{Document Length} & \textbf{gpt-35-turbo} & \textbf{gpt-4} & \textbf{gpt-4o} \\
\midrule
\textbf{Up to 500 characters} & 0.87 & 0.96 & 0.96 \\
\textbf{500-1500 characters} & 0.82 & 0.93 & 0.92 \\
\textbf{1500-2500 characters} & 0.80 & 0.92 & 0.94 \\
\textbf{2500-4000 characters} & 0.70 & 0.88 & 0.91 \\
\textbf{4000-5000 characters} & 0.56 & 0.83 & 0.89 \\
\textbf{5000-6000 characters} & 0.13 & 0.48 & 0.47 \\
\textbf{6000-7000 characters} & 0.08 & 0.14 & 0.33 \\
\bottomrule
\end{tabular}
\label{tab:evaluation-rq3}
\end{table}

As expected, consistency verification becomes more challenging as document length increases. Up to 5000 characters, the gpt-4 and gpt-4o models performed reasonably well. However, beyond 5000 characters, the detection accuracy significantly dropped. Analysis of the results indicates that this decline is due to two primary issues: cases where information is lost during the Markdown conversion process, and cases where inconsistencies are not flagged despite complete data preservation. Both issues tend to occur more frequently with documents exceeding 5000 characters. However, specific characteristics of the design documents that contribute to these issues—such as certain document structures or content types—were not identifiable from the experimental results. Further research is needed to address the limitations observed with longer documents and to explore document characteristics that may impact detection accuracy.

\begin{tcolorbox}[colframe=black,colback=white,boxrule=0.5mm]
\textbf{Answer to RQ3:} The experimental results indicate that detection rates decrease as document length increases, particularly beyond 5000 characters. This suggests that the current method struggles with longer documents. However, the method remains effective for documents under 5000 characters, demonstrating its practical applicability for shorter documents.
\end{tcolorbox}

\section{Implementation and Internal Trial}

Building on the insights gained from our experimental results, we initiated Proof of Concept (PoC) and Proof of Business (PoB) projects across various domains, including finance, automotive, and insurance, to explore the transition from research to practical application. As part of this effort, we deployed a web-based review tool within our organization, designed to allow system engineers to perform reviews directly through their browsers. The demo screen of this tool is shown in Figure \ref{fig9}.

\begin{figure*}[t] \centerline{\includegraphics[width=\linewidth]{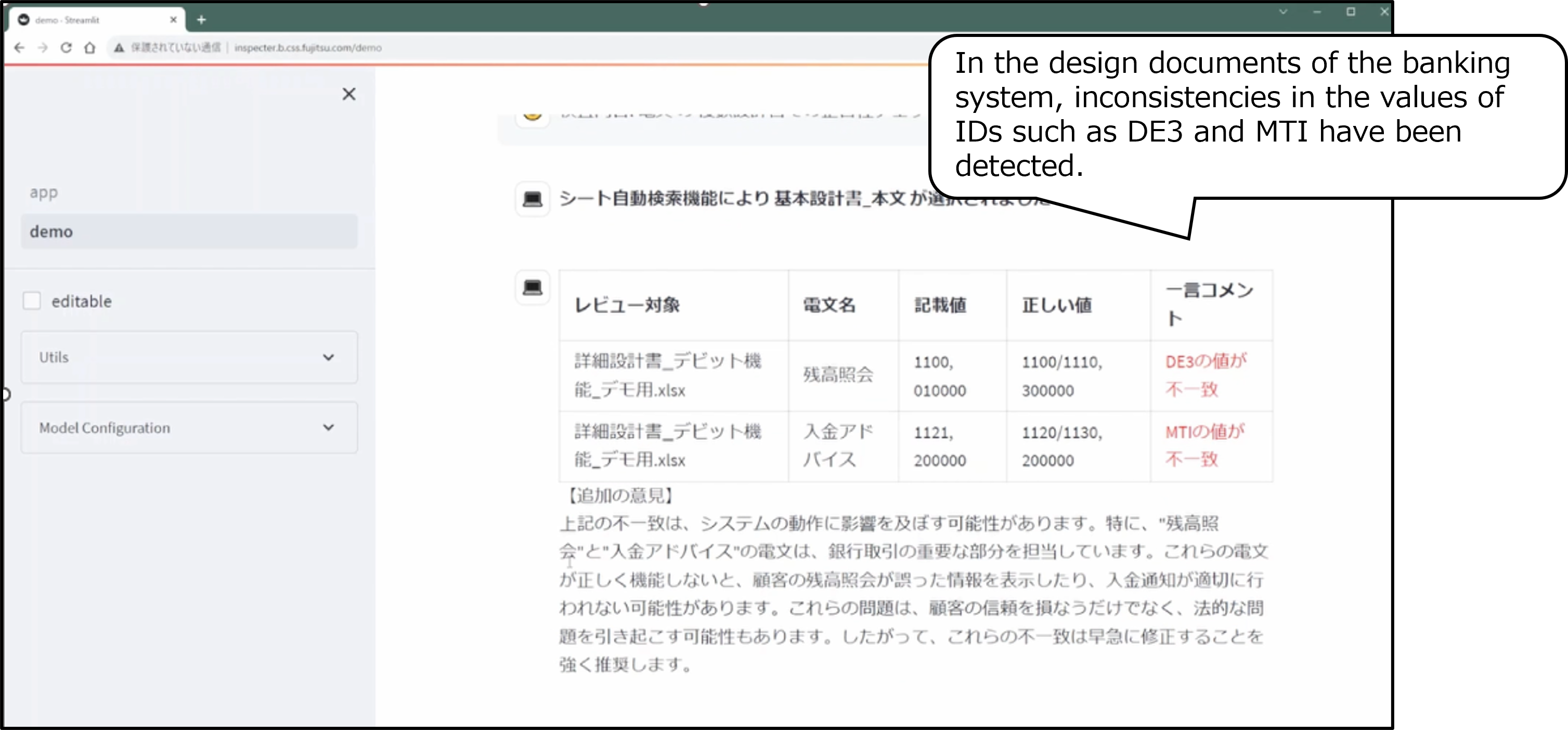}} \caption{Demo Screen of Internal Review Tool.} \label{fig9} \end{figure*}

In the web-based review tool, users start by uploading Excel files they wish to review, which are then automatically converted using the proposed method. Users can create prompts tailored to specific review perspectives and conduct reviews based on these prompts. To assist users, information from Table 1 is provided as a guide, detailing the available review perspectives and their applications. Additionally, the tool includes a catalog of reference prompt information for each perspective, allowing users to quickly generate relevant prompts and streamline the review process.

This tool is actively used within our organization and has been incorporated into several large-scale PoC and PoB projects, involving hundreds of system engineers. Based on PoC results in the automotive domain, nearly 70\% of software design documents exceed 5000 characters, highlighting the need for further technological advancements to accommodate longer documents in real-world applications. However, the current method has demonstrated adequate applicability to approximately 30\% of documents under 5000 characters, suggesting a viable foundation for practical use even at this stage.

\section{Conclusion}

In this study, we aimed to reduce the cost and improve the quality of design document reviews by automating the review process using Large Language Models (LLMs). Initially, we categorized review perspectives based on their difficulty level, distinguishing between areas where general-purpose LLMs can be applied and those requiring specialized LLMs. For review perspectives that do not necessitate specialized LLMs, we proposed a novel method for integrating design documents into LLMs for automated reviews. Our evaluation experiments demonstrated the effectiveness of the proposed method in consistency checks and showed that partial automation of reviews is feasible using existing LLMs. 

While our ultimate goal is to achieve fully automated design document reviews using LLMs, several technical challenges remain. These challenges include:

\begin{itemize}
  \item \textbf{Development of Review-Specific LLMs}: Constructing specialized LLMs to handle the higher-level review perspectives (Levels 3 and 4) among the 11 review perspectives.
  \item \textbf{Development of Optimal Prompting Techniques}: Creating optimal prompting techniques for each review perspective.
  \item \textbf{Scalability for Large Design Documents}: Improving the scalability of LLMs to handle large-scale design documents effectively.
  \item \textbf{Recognition of Image Information}: Enhancing LLMs to understand and interpret image data within design documents. With the recent advancements in multimodal LLMs, there is increasing potential for these models to effectively process and analyze both text and image data in design documents.
  \item \textbf{RAG (Retrieval-Augmented Generation) Techniques}: Developing techniques to automatically and comprehensively identify the necessary design documents for review.
\end{itemize}

Each of these challenges varies in technical difficulty and potential impact, but all are crucial areas of research. We plan to address these challenges in our ongoing research and development efforts.

\appendix
\section{Review Perspectives and Their Overviews}

\renewcommand{\thetable}{A.\arabic{table}}
\setcounter{table}{0}

\begin{table}[h]
\caption{Review Perspectives and Their Description}
\centering
\begin{adjustbox}{center}
\begin{tabular}{|p{2.3cm}|p{5cm}|}
\hline
\textbf{Review Perspective} & \textbf{Description} \\ \hline
Sufficiency Check & Whether it follows the design standards set by the project \\ \hline
Standard/Regulation Check & Whether it follows the development standards/regulations set by the project \\ \hline
Traceability Check & Whether it aligns with the definitions from the upper processes \\ \hline
Compliance Check & Whether it follows the common specifications \\ \hline
Functional Requirement Check & Whether the design content meets the functional requirements \\ \hline
Consistency Check & Whether the content is consistent across design documents \\ \hline
Feasibility Check & Whether it is feasible to implement and maintain the design \\ \hline
Clarity Check & Whether the expressions are clear and understandable \\ \hline
Non-functional Requirement Check & Whether the non-functional requirements are addressed in the design \\ \hline
Cross-sectional Check & Whether the design is consistent across the entire system \\ \hline
Comment Reflection Check & Whether review comments have been correctly incorporated into the design documents \\ \hline
\end{tabular}
\end{adjustbox}
\label{tab:review-perspectives}
\end{table}

\renewcommand{\thefigure}{A.\arabic{figure}}
\setcounter{figure}{0}

\begin{figure*}[t]
\centerline{\includegraphics[width=\linewidth]{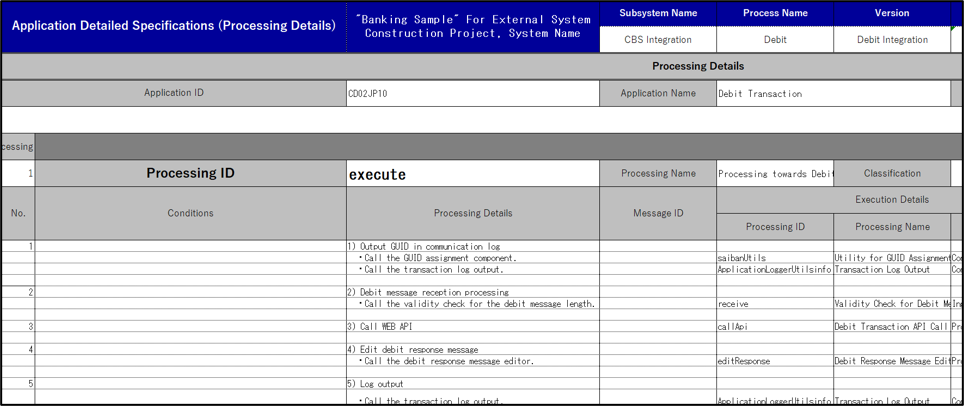}}
\caption{Example of an Excel-Based Design Document for Consistency Check.}
\label{figa1}
\end{figure*}

\begin{figure*}[t]
\centerline{\includegraphics[width=\linewidth]{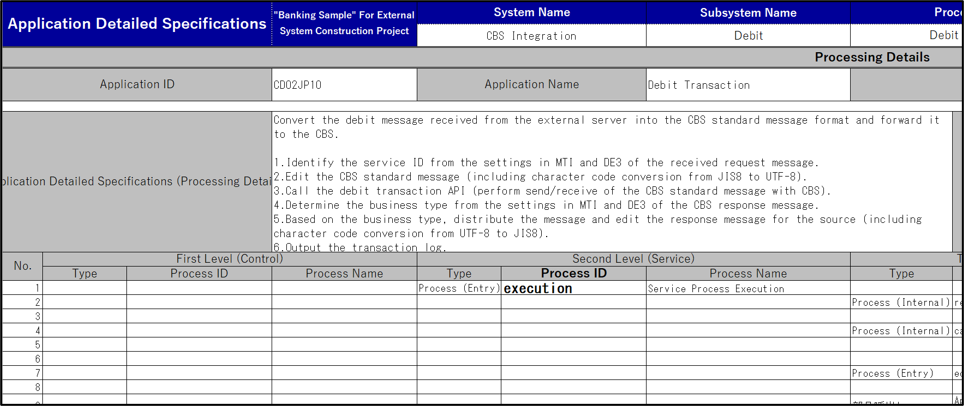}}
\caption{Example of an Excel-Based Design Document with Introduced Inconsistency for Consistency Check.}
\label{figa2}
\end{figure*}

\begin{figure*}[t]
\centerline{\includegraphics[width=\linewidth]{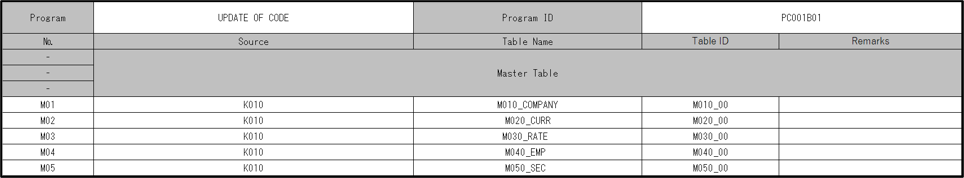}}
\caption{Example of an Excel-Based Design Document with High Frequency of Symbolic Representations.}
\label{figa3}
\end{figure*}

\vspace{12pt}
\end{document}